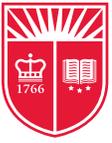

RUTGERS
Edward J. Bloustein School
of Planning and Public Policy
CENTER FOR URBAN POLICY RESEARCH

**Bloustein Local**

# First, Do No Harm:
# Algorithms, AI, and Digital Product Liability

## Managing Algorithmic Harms Though Liability Law and Market Incentives

September 2023

Marc H. Pfeiffer, MPA
Senior Policy Fellow and Assistant Director
Bloustein Local, Center for Urban Policy Research
Edward J. Bloustein School of Planning and Public Policy
Rutgers, The State University of New Jersey

# Bloustein Local and the Center for Urban Policy Research

New Jersey is served by more than 1,500 distinct local government agencies: municipalities, school districts, utilities, counties, and more, all subject to various degrees of state government oversight. **Bloustein Local**, CUPR's local government research unit, has the mission to support **New Jersey state and local government fiscal, administrative, and technology activities.** Our activities include:

- Encouraging and conducting applied and academic research
- Development of resources that can assist others in conducting related research and analysis
- Presenting at mission conferences and symposia
- Contract research and on-call advice for organizations and institutions engaged in mission related activities
- Promote and increase public understanding of issues by partnering with and supporting civic and media organizations that inform and educate the public on state and local government matters.

The **Center for Urban Policy Research (CUPR)** works to make human settlements more equitable, sustainable, resilient, and healthy through research, public engagement, education and other forms of capacity-building.

CUPR is umbrella center that combines the former Bloustein stand-alone Center for Urban Policy Research with the former Rutgers Center for Green Building (RCGB), Environmental Analysis and Communication Group (EAC), Center for Energy Environmental and Economic Policy (CEEEP), and Bloustein Local Government Research Center.

The Center's structure creates strong capabilities for addressing today's challenges in a collaborative and multidisciplinary fashion. These challenges include climate change adaptation and mitigation, coastal hazards, environmental justice, a healthier built environment, sustainable energy transition, and technological transformations affecting daily life and local governance. The Center for Urban Policy Research is recognized for its research on the most critical issues facing communities today including:

- Climate mitigation and adaptation planning
- Land use planning, policy, and practice
- Affordable housing and community-based economic development
- Development, health, and environmental impact analysis
- Energy and environmental policy
- Public finance and government administration
- Historic and cultural preservation
- Infrastructure investment

**https://cupr.rutgers.edu/about/**

# Executive Summary

This report addresses individual and societal harms that result from algorithms that are embedded in digital technology goods and services, products that can create *algorithmic harm(s)*. These products include (but are not limited to) technologies such as generative artificial intelligence chatbots, social media, virtual reality, Internet of Things, robots, etc. This report provides a pathway to reduce algorithmic harms by incentivizing developers to *first, do no harm* as opposed to *work fast and break things*.

- It requires developers to identify and mitigate potential algorithmic harms *before* new products are released and remediate existing ones when harms are discovered. It reflects current trends in cybersecurity, where developers are expected to build security into products and quickly remediate existing products when new risks are found.

- It requires the thoughtful design of definitions of digital products and algorithmic harms. There is a rich trove of academic, non-profit, and corporate research discussing the range of harms. Defined harms must be serious enough to affect the public interest.

- Developers accused of creating unanticipated harm will have a safe harbor if they prove they used contemporary best practices to mitigate any foreseeable potential harms. They would be offered time to remediate them and reduced penalties.

- This process will likely slow development of some digital products. It requires developers to ensure that products are thoroughly tested and that potential adverse outcomes are mitigated *before* deployment. That may delay or limit returns on investment or extend development cycles. In some cases, application creators may decide to abandon products mid-development if harms cannot be sufficiently managed.

Key elements include:

1. Expanding traditional legal liability principles by enhancing legal standards for negligence and product liability to include algorithmic harms:

   a. Negligence liability: expand the requirement of "duty of care" for developers of digital products to include *preventing algorithmic harm* in the product.

   b. Product liability: include algorithmic harm as a type of product defect, injury, or harm.

   c. Reconcile federal standards with state interests.

2. Authorize federal and state regulatory and justice agencies to accept and bring liability complaints of algorithmic harm caused by developer negligence in fulfilling their duty of care by offering defective products.

   a. Permit class actions to be brought by third parties on behalf of groups or society at large.

b.  Provide the judiciary any necessary authority to manage and consolidate like cases. Federal definitions might preempt individual state policies. Provisions may be necessary to allow new laws or regulations to be negotiated and enacted to address potential harms for new products as they develop.

3.  Establish matrices of harms and penalties that address the range of harms, from incidental to substantial and from individual to societal. At the extreme end of substantial societal harm, they must be financially significant enough to discourage undue liability risk-taking.

4.  Developers would be incentivized by their liability insurers to engage in harm prevention during development and deployment. Liability insurers would require that sound harm mitigation standards be met to secure and maintain coverage.

This framework requires technology policy and legal subject matter experts to elaborate on and refine the details. Input, balance, and compromise from societal, financial, and technological interests are are at the core of its potential.

Additionally, these ideas do not have to stand alone. They can be integrated into other solutions being discussed. This is particularly important as these issues are currently top of mind for many federal and state lawmakers.

While focused on the United States and its liability practices, the model will likely have value in other countries if adapted to local circumstances.

Society needs sound algorithmic focused public policies that incentivizes harm prevention.

We should stop breaking things.

# Preface

A new breed of algorithmic decision-making is spreading like a virtual wildfire, disrupting broad swaths of society, culture, health, commerce, and physical environment with the potential for a conflagration.

Arguably, we are at a societal and economic inflection point. Unlike preceding innovations such as the printing press, electricity, and of course, the internet itself, this one is producing a greater impact in a much shorter time.

Algorithm-powered digital technology is moving at a pace that is evolving and growing day-by-day. It is exposing a reality that humans have limits in how quickly economic, socio-cultural, and political change can be successfully absorbed.

Financial, academic, technological, and social media sectors, among others, are looking to manage the disruptions with different approaches depending on their vested interests. These solutions run a gamut from traditional to innovative, complementary to adversarial, and focused to comprehensive.

Controlling the wildfire and the damage it creates is proving difficult to do. But how can we go about preventing harm in the first place? This paper is intended to do just that.

It is offered to enhance existing proposals for digital technology regulation. These proposals cover a wide range of risk management and regulatory approaches. Government administrations and legislatures in the US and around the world are studying these alternatives to identify, analyze, manage, and monitor the risks that AI poses. It is hoped these ideas will provide a firebreak to dampen the wildfire.

The ideas and opinions expressed in this report are those of the author. The report is not a traditional academic paper in research approach or style. The ideas are informed by his experience over the last 45+ years dealing with digital technology and risk management in public organizations. He avidly follows contemporary reporting and dialog on these issues. They do not necessarily represent the view of Rutgers University or any of its affiliates or agencies.

The author also thanks his editor Debra C. Meltzer for her many invaluable contributions to the readability and style of this report. Karyn Olsen of the Bloustein School is also thanked for her preparation of the formal report document.

# Introduction

Lawmakers and policy advocates have proposed the creation of a federal government regulatory environment in which algorithmic-based digital applications would be submitted to an agency for "approval." However, the idea that applications using artificial intelligence innovations would be required to meet a set of standards has unforeseen and irresolvable technical and legal challenges. While well intended, they are likely destined to fail.

Technological innovations move much more quickly than government policy development. The time it takes to establish licensing, regulatory, and permitting procedures will inevitably slow innovation and its potential economic and societal benefits to am unnecessary  crawl. A new federal agency needs time to coalesce; the time frame for staffing and establishing practices is a speculative and time-consuming undertaking. That is particularly true in a time of low unemployment coupled with a high demand for technology and specialized policy expertise. It carries a high risk of failure, regardless of the proposal's aspirational words and intent. Adding to the portfolio of existing agencies is less risky, but not without similar challenges and the added risk of inter-agency conflicts.

Yet, on behalf of civil society, governments have a responsibility to manage the risks and harms inherent to digital technologies that use algorithms to make decisions without human intervention. These are found in AI and other existing and yet to be created digital products (hereafter, *product[s]*). They present societal risks that must be managed without compromising underlying constitutional principles (e.g., those of due process and the freedoms of speech and of the press).  Likewise, they need to avoid hamstringing innovation and the role of the market.

A practical approach would incentivize investors and developers to switch from "*move fast and break things*" to "*first, do no harm*." Absent societal guardrails, Mark Zuckerberg's motto undermines safety and ignores harms. Preventing algorithmic harm (or hereafter, *harm[s]*) happens only after damage is done because there is no incentive to spend development resources on mitigating them until a complaint has been lodged.

On the other hand, Hippocrates' harm-avoidance approach has worked reasonably well for medicine, engineering, and other practices for generations. Additionally, we need to appreciate that regulation that foresees every possible product and risk is not attainable; the perfect cannot be the enemy of the good and uncertainties cannot be avoided. We need to develop procedures to manage them.

# The Problem

We live in a society where there are too many individuals and organizations, and nation states that misuse digital technology for dishonest purposes. They act maliciously to steal data or manipulate users through false advertising, fake evaluations and product reviews, and the perpetration of out-and-out fraud. Their goals are to monetize their efforts or disrupt society. They are often motivated by the rigors of the marketplace where competition can force players to cut corners or take undue risks with their products. A quick scan of news reports can produce any number of examples of pernicious behavior by individuals and organizations.

Unfortunately, some developers of digital products engage in this type of unethical behavior as well. They purposely design their applications to surreptitiously leverage user data and illegally influence their position in the competitive marketplace. Many organizations delay or refuse to respond to these threats until the public outcry becomes too great, and then their remedies may be marginal and insufficient to neutralize the constantly evolving threats.

Regrettably, civil society needs to acknowledge these realities and work to protect unsuspecting users, organizations, and the public from harm. It must ensure that innovations that reach the market have built-in safeguards. Today, technology ethicists and experts around the world are studying a wide range of policy and political approaches to control the different forms these disruptions can take.

This proposal is intended to contribute to the ongoing discussions in civil society as it attempts to manage the problem of algorithmic harms. It does so by using market forces, government guardrails, and principles of liability law.

---

**Definition of Algorithmic Harm***

*Algorithmic harms refer to the unintended negative effects, consequences, or biases that arise from the design, implementation, or execution of algorithms and automated processes in various technological systems. These harms can manifest in a range of contexts, including artificial intelligence (AI), machine learning, data analytics, and decision-making algorithms.*

**Details and bibliography in Appendices A–C**

*Courtesy of ChatGPT prompt to "Define 'algorithmic harm' and reviewed by the author.*

---

# Why Liability Law?

Enhancements to current U.S. liability laws to address algorithmic harms would force developers to consider and manage the full range of potential risks engendered by their products as a standard practice. This approach leverages market-based incentives with an expanded set of liability laws serving as guardrails. Together they would address the harm that poorly designed algorithmic-based systems could bring.

The concept of using liability guardrails is not original.[1]  A recent policy brief by the Centre for International Governance Innovation, Addressing the Liability Gap in AI Accidents[2] by Amrita Vasudevan, is an excellent assessment of the issues and challenges, with extensive references on the issue. This proposal expands on the concepts in this and other papers. Elements of the proposal were recently addressed in a joint letter to President Biden and his technology

---

[1]  While not used in the development of this paper, other authors have addressed the general concept of product liability in AI. A representative sample are listed in Appendix E.

[2]  https://www.cigionline.org/static/documents/PB_no.177.pdf

leadership team from a group of advocacy organization coordinated by Upturn.org.[3]

These kinds of harms are treated as economic externalities in our investment and economic systems. They do not provide necessary incentives or signals for developers to prevent them.  Potential financial returns drive product design, investment, and marketing decisions. As part of its societal role, government must step in to prevent societal and individual harms from algorithms. Market forces on their own do not provide sufficient incentives to do that.

A prime example of this is Section 230 of the Communications Decency Act enacted in 1996.[4] The law created a virtual technological autobahn, the impact of which was not foreseen at the time. Think of 230 as a wide highway without traffic controls, sign guidance, lane markings, medians, safety features, exit ramps, speed restrictions, or laws governing driver and pedestrian safety. Not a guardrail in sight.

Government had no reason to assess future risks when 230 was enacted. The internet was new, its potential clouded. In retrospect, we now realize an opportunity was missed. Adding liability for algorithmic harm to our legal arsenal can establish balance by giving all parties the incentive to act responsibly. Carefully crafted, government liability guardrails do not need to inhibit freedom of speech or the press, nor does it need to hamper economic innovation.

When harm comes up against free speech, it is not the speech being regulated, it is the behaviors and decisions that lead to or follow the speech that are being regulated. Many tools and practices to moderate speech already exist, and others can be designed to prevent or mitigate the potential for harm, without compromising developers' right to choose the services they provide. Being exposed to well-constructed liability penalties for failure to protect will encourage informed and cautious development decisions.

Using algorithmic risk as a basis for liability affects a wide range of digital technologies. Beyond generative AI and its cousins, this encompasses social media, surveillance tech, health applications, virtual reality, autonomous vehicles, the Internet of Things, and technologies yet to come. Algorithmic AI is becoming ubiquitous and will be integrated into other technologies over the coming decade as another tool. It is the harm that algorithms can bring that must be addressed, not any single form of digital technology.

# Harms, Risk, and Liability Practices

Liability risk incentivizes developers and their investors to manage exposure at the start of the development cycle. It jeopardizes funding if technologists and investors fail to consider liability risk during the development process. Over the last 15 years, digital developers have learned the importance of including cybersecurity protections at the beginning and throughout the life cycle of their development practices. History teaches us that fixing things after they break is not a smart way for societies to behave. Fixing things after the fact costs more than doing it at the start.

This proposal suggests the enhancement of existing liability laws to include algorithmic harms. It would impose financial penalties tied to the degree of damage relative to the scope of a harm caused by a product that failed to competently assess, manage, and mitigate algorithmic harms *before* a product goes into public testing or use.

This approach would have prevented the well-known issues that ChatGPT experienced in its initial "public testing" release in November 2022.[5] Merely posting a notice that "this is a test and there may be flaws" is arguably inadequate given that human nature tends to run and push limits on something new. Flaws should be discovered in a controlled environment, not in a public and media free-for-all. Meta recently used this approach in deploying chatbots as open source with similar effects.[6]

If products can be demonstrated to create significant algorithmic harm, they should be subject to liability law litigation. Development activities must address potential harms using industry standard practices with (to be developed) documentation and frameworks to support them. If the complaint review finds that a documented development process was sound, it could grant the product limited safe harbor protections and time to remediate harm, thus reducing monetary and reputational risks.

This approach does not require licensing, standards, or auditing approaches, though third-party audit tools and credible development frameworks will likely become part of the development toolbox.

Regrettably, existing liability theory and law have not evolved to address the complexities that algorithmic-based products bring. To bring this about, a supplemental legal regime will need to enhance negligence and product liability practices, specifically:

- Negligence liability: expanding the *duty of care* principle. This would require developers of digital products (goods and services) to foresee and prevent harms caused by digital technology that includes algorithms in the product.

- Product liability: include algorithmic harms as a type of product defect, injury, or harm.

In addition, it must address:

- Definition of algorithmic harms: The definition must focus on the impact of harms to individuals, groups, and society-at-large. Academic experts have written widely on the impact and implications of algorithmic harms that can affect the public interest. To elaborate on this point, the author queried several generative AI chatbots for an in-depth description of algorithmic harm and a bibliography of verified references. After ensuring they are all real sources, most results appear valid and are included as Appendix A. Two additional perspectives are provided. The Electronic Privacy Information Center (EPIC) highlighted AI harms in its report, [Generating Harms: Generative AI's Impact & Paths Forward](#)[7] (summary in Appendix B). A separate chatbot query for "general societal harms" produced the thoughtful list in Appendix C.

5  One example of many includes [www.bleepingcomputer.com/news/technology/openais-new-chatgpt-bot-10-dangerous-things-its-capable-of/](http://www.bleepingcomputer.com/news/technology/openais-new-chatgpt-bot-10-dangerous-things-its-capable-of/)

6  As one example, see What Can You Do When A.I. Lies About You? New York Times, August 3, 2023, [www.nytimes.com/2023/08/03/business/media/ai-defamation-lies-accuracy.html](http://www.nytimes.com/2023/08/03/business/media/ai-defamation-lies-accuracy.html)

7  [https://epic.org/documents/generating-harms-generative-ais-impact-paths-forward/](https://epic.org/documents/generating-harms-generative-ais-impact-paths-forward/)

At the same time, the definitions of harms need to exclude incidental or outlier annoyances and focus on individual, group, and systemic risks that are substantial in nature. This leads to a process that should include other elements.

- Developing a combination of administrative and judicial dispute resolution procedures, particularly in the context of the roles played by state and federal administrative agencies and judiciaries in the process. In addition to traditional court litigation, alternative dispute resolution procedures will be important.

- The traditional standard of proof for liability cases, i.e., a preponderance of the evidence, could apply in cases where remediation is warranted. For cases where developers acted without regard to harm and substantial financial penalties are justifiable, the higher standards of clear and convincing evidence may be warranted.

- General guidance available for developers on technological mitigation practices. It may be instructive to reflect on the lessons learned from the time when cybersecurity started as a programing afterthought; current best practices include integrating it into developing products that are secure by design and secure by default. Effectively done, it will incentivize the development of new tools, perhaps even market-driven competition to develop new ones, that will raise the protection bar for all over time. There are already government and non-government organizations that have the capacity to and are already developing robust mitigation models and tools.

- Developers need to understand the risks of their development process, consider potential unanticipated consequences, and integrate ways to mitigate them.

- Consideration of the traditional roles of federal standards and state prerogatives of liability laws, reconsidered for this new risk.

This approach will likely establish new and necessary precedents. It will draw from the rich history of liability laws dealing with consumer protections, food and drug safety, financial market investments, occupational safety, environmental hazards, etc. While the range of algorithmic harms may be similar in some ways, it is different from what came before it. Legal and technological expertise will be needed to translate this proposal into specific and actionable policies.

Developers that fail to manage the risks in advance of a product roll-out or fail to monitor existing products for harm, will face the risk of liability complaints. If what they release results in algorithmic harms that could or should have been prevented, they should be subject to substantial financial or operational penalties.

This approach has been effective with the European Union's General Data Protection Regulation (GDPR) and the promising Digital Services Act where violators can face severe financial penalties. It has already been shown that the potential of these penalties has altered corporate behavior (e.g., disclosure and limits on information gathering cookies).

In the event a developer is accused creating a foreseeable harm, a complaint resolution process should be initiated. This process could clear the developer or reduce a penalty if it were shown that the company engaged in prudent actions to identify and mitigate harms or that a particular harm could not have been reasonably foreseen. This is the concept of a safe harbor.

In such cases, the developer would be given time to fix the flaws or withdraw the product from the market and face limited financial exposure. Examples include auto and consumer product recalls, regulation of financial products and the developing field of cybersecurity vulnerabilities. If no action is taken, the developer could face more substantial penalties.

The adjudication process should also build in opportunities for state and local governments to develop new policies as needed to address algorithm-driven risks along with new and potentially useful innovations before they are deployed. Examples of this include the air quality effects and congestion on cities, and with labor markets of ride-sharing and short-term housing rental, aka, Uber/Lyft and Airbnb/VRBO.

# Mitigation Tools

There are many ways developers can mitigate individual and societal harms. Those can include user-facing tools such as clear and plain language notices, credible and understandable opt-in and opt-out alternatives, check boxes, plain language disclosures of potential harms, red flag warnings, avoidance of dark patterns,[8]  know your customer (KYC) practices, and more.  They can also conclude that an idea has too much risk or potential harm to move forward with its development.

Application development mitigation practices can also include third-party audits of products, validated training data and related inputs, documentation of development risk mitigation decisions, bias assessments, independent evaluations of worst-case scenarios, and more. These tools can help determine and avoid potential harms. It is also likely that risk management and mitigation framework methodologies can be developed by organizations like NIST, CISA, ISACA, CIS, ISO, and various ISACs.[9]

In some cases, simply and clearly informing users of the technological risks and having them "check the box" to accept them may be adequate. But there must be a balance between reasonable risks that users can manage versus ones that must be eliminated at the start. Those decisions may end up serving as the basis of an evolving "reasonable developer" standard.

Insurance is another risk management tool that enables organizations to protect financial and intellectual assets and mitigate risks. Once algorithmic harm liability is defined, new insurance markets will evolve to insure developers against vulnerabilities. Business liability insurers would then sell insurance products at various levels of insurance coverage.

Insurers often protect themselves against the risks presented by their clients by imposing exposure mitigation requirements on them. In this case, insurers will likely require the client to document their research into potential harms and show evidence of the steps they take to mitigate them (i.e., not using incomplete or biased training data or poorly tested algorithms) before launching products. If they fail to meet the expected standards, the insured developer will risk being denied coverage if a claim is made.

---

8  https://en.wikipedia.org/wiki/Dark_pattern
9  See Appendix D for explanation of these acronyms.

This happened in cybersecurity. As cyber risks evolved over the last five years, insurance policies met the challenge by requiring clients to meet minimum best practice security standards as terms of coverage. This has had a major impact on reducing cybercrime from what it might have otherwise been.

# Conclusion

This embryo of a proposal incentivizes investors and developers to ensure that potential algorithmic harms are addressed before they occur in new digital products. The intent is to make developers attentive to their risks and more cautious about how and what they develop and deploy. That may result in slowing down the pace of development or innovation. That is not necessarily a bad reaction, as a more thoughtful approach could substantially lower long-term risks.

It also requires governments to adopt policies that incentivizes reduction of the adverse impacts of new products. Using a well-developed liability framework would shift the responsibility for harm protection to the market and away from overt government regulation, but use informed guidance and societal judicial processes to ensure compliance.

Fleshing this out requires subject matter experts in harms and liability law to move the theory through a thicket of legal issues, political obstacles, and the potential for unanticipated consequences to reach the goal. The goal is embedding risk management of harms and legal liability into the development and distribution of algorithmic-driven digital products. It should shift the investment emphasis from moving fast to addressing the harmful impacts of products.

In the long term, if the model is successful and a critical mass of governments adopt the model, it could reduce harm at an international level and minimize the impact of outliers who choose to circumvent mainstream practices.

This use of product liability as public policy should be part of the current dialog and debate about AI and algorithms. Other ideas may complement, integrate, or enhance this approach. But the notion of combining economic disincentives with smart guardrails, defined harms, and a model to resolve complaints may be a way through the challenges and opportunities that algorithmic decision-making presents to society.

In the end, should society let developers deploy algorithms without taking responsibility for them, just because they can?

No, because it will break things.

# Appendix A - What is an Algorithmic Harm? And a Bibliography

*Response of ChatGPT to the prompt: Define "algorithmic harms"*

**Algorithmic Harms: Unintended Consequences of Automated Processes**

**Definition:** Algorithmic harms refer to the unintended negative effects, consequences, or biases that arise from the design, implementation, or execution of algorithms and automated processes in various technological systems. These harms can manifest in a range of contexts, including artificial intelligence (AI), machine learning, data analytics, and decision-making algorithms.

**Characteristics:**

1. Bias and Discrimination: Algorithms can unintentionally perpetuate or amplify biases present in training data, leading to discriminatory outcomes against certain groups.

2. Privacy Violations: Improperly designed algorithms might extract sensitive information or infringe upon individuals' privacy rights.

3. Unfair or Unethical Outcomes: Algorithms can generate results that are unfair, unjust, or morally problematic, often due to flawed assumptions or inadequate modeling.

4. Misinformation Spread: Automated content recommendation systems may inadvertently promote or amplify false information, contributing to misinformation spread.

5. Loss of Human Autonomy: Overreliance on algorithms can diminish human decision-making agency and control, leading to potential disempowerment.

6. Economic and Social Disparities: Algorithmic systems can exacerbate socioeconomic inequalities, impacting access to opportunities and resources.

7. Security Risks: Poorly designed algorithms might create vulnerabilities that malicious actors can exploit for cyberattacks or unauthorized access.

8. Health and Safety Concerns: In sectors like healthcare or autonomous vehicles, algorithmic errors can have life-threatening consequences.

---

**Examples:**

1. Criminal Justice Algorithms: Predictive policing algorithms may disproportionately target certain communities, leading to biased law enforcement.

2. Social Media Algorithms: Content recommendation algorithms can reinforce echo chambers and promote polarizing or harmful content.

<h1 style="text-align:center;color:red;">Appendix A (continued</h1>

3. Credit Scoring Algorithms: Biases in credit assessment algorithms might result in unjust credit denials for marginalized groups.

4. Autonomous Vehicles: Errors in self-driving car algorithms could lead to accidents with human and property casualties.

**Addressing Algorithmic Harms:**

1. Ethical Design: Algorithms should be developed with ethical considerations, transparency, and accountability in mind.

2. Diverse Data: Ensure training data is representative and diverse to mitigate biases.

3. Ongoing Monitoring: Continuously assess algorithmic outcomes for unintended consequences and biases.

4. Regulation and Oversight: Establish regulations and frameworks to hold companies accountable for algorithmic harms.

5. Public Awareness: Raise awareness about algorithmic impacts among users, policymakers, and stakeholders.

Algorithmic harms underscore the importance of responsible development, rigorous testing, and ongoing evaluation to minimize unintended negative effects and ensure the positive impact of algorithmic technologies on society.

### Algorithmic Harms Bibliography

As a chatbot exercise, the author prompted the Claude.AI chatbot for the following:

*"Can you develop a bibliography and provide urls to 15 of the most important papers and books that discuss algorithmic harms. You cannot invent any research, only use those found on web."*

The result from Claude included a disclaimer that reads "*Links provided by Claude may not always be valid or up to date.*" Claude was prescient, its list had flaws.

- Eight references were legitimate.  Of them, six had bad links, but were discoverable via a title search. The correct links are shown below.

- Two references did not exist.

- Three items were duplicates (listed twice).

- One reference sounded good, but the link required additional clicks to find content, and one was not available to the public.

# Appendix A (continued

The eight useful links are shown below with the original numbering sequence provided by Claude and without the flawed items. The flawed items follow that list.

**Response of Claude.AI**

Here is a bibliography with URLs to 15 important papers and books discussing algorithmic harms:

1. Dissecting Racial Bias in an Algorithm that Guides Health Decisions for 70 Million People (2019)
https://dl.acm.org/doi/10.1145/3287560.3287593

2. Discrimination, Artificial Intelligence, and Algorithmic Decision-Making (2018)
https://rm.coe.int/discrimination-artificial-intelligence-and-algorithmic-decision-making/1680925d73

3. Algorithms of Oppression: How Search Engines Reinforce Racism (2018)
https://nyupress.org/9781479837243/algorithms-of-oppression/

5. Algorithmic bias: on the implicit biases of social technology (2021)
https://link.springer.com/article/10.1007/s11229-020-02696-y

7. Automating Inequality: How High-Tech Tools Profile, Police, and Punish the Poor (2017)
https://us.macmillan.com/books/9781250074317/automatinginequality

8. Automating Society Report 2020 (2020)
https://automatingsociety.algorithmwatch.org

12. Ruha Benjamin: Race After Technology: Abolitionist Tools for the New Jim Code (2019)
https://ruhabenjamin.com/race-after-technology

14. AI Now Institute: Discriminating Systems: Gender, Race and Power in AI (2019)
https://ainowinstitute.org/publication/discriminating-systems-gender-race-and-power-in-ai-2

*Links provided by Claude may not always be valid or up to date. We appreciate your patience as we work to improve link accuracy*

**Non-existent references**

4. Secret Biases in Algorithmic Hiring: Automating Inequality in the Labor Market (2022)
https://link.springer.com/article/10.1007/s43681-022-00147-5

9. The Guardian: How algorithms can create inequality in health care, finance and other areas (2021)
https://www.theguardian.com/technology/2021/aug/15/how-algorithms-can-create-inequality-in-health-care-finance-other-sector

# Appendix A (continued)

**Duplicated within:**

6. Discrimination, artificial intelligence, and algorithmic decision-making (2018)
https://www.tandfonline.com/doi/abs/10.1080/13600869.2018.1475959

13. Safiya Umoja Noble: Algorithms of Oppression (2018)
https://nyupress.org/9781479837243/algorithms-of-oppression/

15. Virginia Eubanks: Automating Inequality (2018)
https://us.macmillan.com/books/9781250074312

**Poor link**

10. FAT* Conference on Fairness, Accountability, and Transparency (Conference site with published papers)
https://facctconference.org/

**Not accessible to public**

11. Arvind Narayanan: How to recognize AI snake oil (2019)
https://www.youtube.com/watch?v=6vfSFXKlnO0

# Appendix B – Common AI Harms as Described by EPIC[10]

1. **Physical Harms:** These are harms that lead to bodily injury or death, which may include acts by AI companies that facilitate or encourage physical assault.

2. **Economic Harms:** These are harms that cause monetary losses or decrease the value of something, which may include the harms of fraudulent transactions conducted by those using AI to impersonate a victim.

3. **Reputational Harms**: These harms involve injuries to someone's reputation within their community, which may in turn result in lost business or social pariahdom.

4. **Psychological Harms:** These harms include a variety of negative—and legally cognizable—mental responses, such as anxiety, anguish, concern, irritation, disruption, or aggravation. Danielle Citron and Daniel Solove place these harms within two categories: emotional distress or disturbance.

5. **Autonomy Harms:** These harms restrict, undermine, or otherwise influence people's choices and include acts like coercion, manipulation, failing to inform someone, acting in ways that undermine a user's choices, and inhibiting lawful behavior.

6. **Discrimination Harms:** These are harms that entrench or exacerbate inequality in ways that disadvantage certain people based on their demographics, characteristics, or affiliations. Discrimination harms often lead to other types of AI harms.

7. **Relationship Harms:** These harms involve damaging personal or professional relationships in ways that negatively impact one's health, wellbeing, or functioning in society. Often, these harms damage relationships by degrading trust or damaging social boundaries.

8. **Loss of Opportunity:** Related to economic, reputational, discrimination, and relationship harms, loss of opportunity is an especially common AI harm in which AI-mediated content or decisions serve as a barrier to individuals accessing employment, government benefits, housing, and educational opportunities.

9. **Social Stigmatization and Dignitary Harms:** Related to reputational, discrimination, and relationship harms, these harms undermine individuals' sense of self and dignity through, e.g., loss of liberty, increased surveillance, stereotype reinforcement, or other negative impacts on one's dignity.

# Appendix C - List of General Harms Created by Digital Products

*Provided by Claude.AI*

**Prompt:** *Similar to the notions of individual or societal harms that AI systems can cause, what kinds of harms can be created by digital products generally? Provide a bullet pointed list of each one, with the harm in bold face, followed by an explanation.*

**Response:**

Here is a bullet pointed list of potential harms caused by digital products:

- **Privacy violations** - Many digital products collect large amounts of user data, and if not handled properly, this data could be misused or expose users to privacy risks. Lack of transparency around data collection and use can also erode user trust.

- **Addictiveness** - Some digital products, especially social media apps, are designed to be very addictive and encourage continual use. This can have negative impacts on mental health, productivity, and work/life balance.

- **Misinformation** - The speed and reach of the internet allows misinformation to spread rapidly on digital platforms. This can lead to real-world harms if people are misled on important topics.

- **Discrimination** - There are risks of biases being embedded in algorithms that lead to discriminatory treatment or impacts on underrepresented groups. For example, biased data used to train AI systems.

- **Economic disruption and inequality** - Digital disruption has impacted many industries and jobs. While innovation can boost productivity, it can also contribute to income inequality if benefits are concentrated among a limited population.

- **Harm to social skills and relationships** - Excessive use of technology and digital products may impair the development of face-to-face interaction and relationships. This is especially a concern for child development.

- **Distraction and reduced productivity** - Digital products compete for our attention and can cause distraction, reduced focus, and decreased productivity. Features to maximize engagement encourage continual scrolling and checking of apps.

# Appendix D - List of Organization Acronyms

EPIC – Electronic Privacy Information Center   www.epic.org

NIST – US National Institute for Standards and Technology  www.nist.gov/artificial-intelligence

CISA – US Cybersecurity and Infrastructure Security Agency www.cisa.gov

ISACA - www.isaca.org

CIS – Center for Information Security www.cisecurity.org

ISO – International Standards Organization  www.iso.org

ISAC – Information Sharing and Analysis Centers
https://en.wikipedia.org/wiki/Information_Sharing_and_Analysis_Center

# Appendix E - A Sampling of References Addressing Liability and Digital Products

Products liability law as a way to address AI harms, John Villasenor, Brookings Institute, October 31, 2019. https://www.brookings.edu/articles/products-liability-law-as-a-way-to-address-ai-harms/

Liability for Artificial Intelligence and other emerging digital technologies, European Commission, 2019. www.europarl.europa.eu/meetdocs/2014_2019/plmrep/COMMITTEES/JURI/DV/2020/01-09/AI-report_EN.pdf

When Algorithms Harm Us, Carrie Kirby, Iowa Law Magazine, November 30, 2022. https://law.uiowa.edu/iowa-law-magazine/news/2022/11/when-algorithms-harm-us

Artificial intelligence and civil liability—do we need a new regime?, Baris Soyer and  Andrew Tettenborn, International Journal of Law and Information Technology, Volume 30, Issue 4, Winter 2022 https://academic.oup.com/ijlit/article/30/4/385/7039697

# About the Author

**Marc H. Pfeiffer, MPA**
Senior Policy Fellow and Assistant Director
Bloustein Local, Center for Urban Policy Research
Bloustein School of Planning and Public Policy
Rutgers University
https://cupr.rutgers.edu/blousteinlocal/
marc.pfeiffer@rutgers.edu

Marc H. Pfeiffer "retired" in 2012 from a 37-year career in New Jersey local government administration, having served as a municipal administrator in several municipalities, and 26 years of service in the state's local government oversight agency, the Division of Local Government Services.  At DLGS he served as Deputy Director for 14 years, and periodically as Acting Director.

Marc has broad experience in many areas of local government policy and administration, including specific expertise in areas such as finance and property taxation, public procurement, shared services and consolidation, technology, energy, labor relations, and general government administration. He also has deep experience in the legislative process and as a regulatory officer. He is currently engaged in research concerning the use of technology in local government.

In addition to participating in Bloustein Local, Marc makes his extensive government experience available as a guest lecturer and other collaborative efforts. He is also assisting the Rutgers School of Public Affairs and Administration with the State's Certified Public Manager Program in curriculum development and instruction.

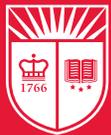

RUTGERS

Edward J. Bloustein School
of Planning and Public Policy
CENTER FOR URBAN POLICY RESEARCH